# Proposed Challenges and Areas of Concern in Operating System Research and Development

Plawan Kumar Rath[1], Anil G.N[2].

[1] Department of Computer Science and Engineering, BMS Institute of Technology
Bangalore, Karnataka, India
plawanrath@gmail.com

[2] Department of Computer Science and Engineering, BMS Institute of Technology
Bangalore, Karnataka, India
anilgrama@yahoo.co.in

## ABSTRACT

Computers are a very important part of our lives and the major reason why they have been such a success is because of the excellent graphical operating systems that run on these powerful machines. As the computer hardware is becoming more and more powerful, it is also vital to keep the software updated in order to utilize the hardware of the system efficiently and make it faster and smarter. This paper highlights some core issues that if dealt with in the operating system level would make use of the full potential of the computer hardware and provide an excellent user experience.

## 1. INTRODUCTION

Computer technology has made incredible progress in the roughly 60 years since the first general purpose electronic computer was created. For the evolution of computers from being just a scientific tool to being a necessity in every household the operating systems that run on them have played a very vital role. Today we don't call a computer system by the manufacturer names; we rather call a system to be a Mac PC or a Windows PC, etc. Although the operating systems are becoming more and more dynamic and classy yet there remains a lot of work to make them utilize the full functionalities of the fast computer hardwares of today. Here we will see some of the key issues that the operating systems face and the unconquered challenges that still remain in the world of operating system research and development. We divide the rest of the paper into four segments. In the first segment we talk about security, in the next we talk about memory management, then we see multiprocessor programming in operating systems and related issues, and finally we will shift our focus onto the smart devices and see the issues in user interface designs for the same.

## 2. SECURITY

Security has been and still remains a major concern for operating system developers and users alike. Informally speaking, security is, keeping unauthorized entities from doing things you don't want them to do. Operating system protection involves protection against unauthorized users as well as protection of file systems. File permissions are based on user identity, which in turn are based on user identity, which in turn are based on authentication. Hence authentication of users has to be highly secure such that any unauthorized user doesn't hack in along with proper mechanism to let in genuine users. Various authentication mechanisms have been and are being used in operating systems, like the old-fashioned password authentication, where a plaintext password is stored. This mechanism has been proven to be easily hackable, so another technique that provides an alternative is Hashed Passwords.

*General Algorithm:*

- Store $f(Pw)$, where $f$ is not invertible
- When user enters $Pw$, calculate $f(Pw)$ and compare

Attackers can still use password-guessing algorithms; therefore most operating systems use access control mechanisms to protect the hashed passwords. Another authentication mechanism used is the Challenge/Response Authentication. Here what happens is the server knows $Pw$ and sends a random number $N$, both sides then calculate $f(Pw,N)$ where $f$ is some encryption algorithm. Although it must be noted that this mechanism is not very famous with operating systems. The reason being that, even in this case a person who guesses $N$ or finds it out and comes to know $f(Pw,N)$ can run password-guessing algorithms, so it is not that very different from the hashed-password authentication in terms of security. These days use of biometrics has become a major user authentication mechanism. Such techniques include fingerprint readers, iris scanner, etc. Although biometrics works fine if used locally, yet even these methods are susceptible to spoofing attacks. Hence we can infer that even the best and the most hi-tech authentication has its limitations.

When talking about operating system security, authentication attacks will make the bottom of priority list. The major problems are attacks like, Trojan Horses, Login spoofing and Buggy Software. Trojan Horses are basically programs that are disguised programs, meant to harm the system and its resources. Someone may be tricked into running a program that may adversely affect that user; his system or data. Although Linux, UNIX and other Unix-like operating systems are generally regarded as very protected, yet they are not immune to computer viruses. For example, consider a virus program written in C, which goes on creating new files and allocating space in an infinite loop! Will Linux be safe in that case? Hence viruses are a threat to all operating systems. Although it must be noted that there has not yet been a widespread linux malware (malware as in any malicious software) threat of the type that Microsoft Windows softwares face; this is mostly because of the following reasons:

- The user base of the Linux operating system is smaller compared to Windows.
- The malwares' lack root access.
- Fast updates for most Linux vulnerabilities.

Operating systems may use the following mechanisms to avoid attacks of this type:

- Operating Systems can provide sandboxes: Sandboxes are environments where a program can execute but should not affect the rest of the machine.
- The trick here is, permitting limited interaction with outside while still providing the full functionality of the operating system. Or in other words the file system can be kept out of unauthorized access and $3^{rd}$ Party softwares may be allowed minimum access to file-systems.

Race conditions can also be a critical security issue. To illustrate such a situation, consider a privileged program that checks if a file is readable and then tries to open it as root. The attacker passes it a symbolic link, in the interval between the two operations; the attacker removes the link and replaces it with a link to a protected file. This would give him direct access to the protected file area and into the system. So here an attacker takes advantage of the race condition between two operations to get access into the protected area of the operating system. The only way to overcome such attacks is to provide only atomic operations to access files and strict restrictions on their access by other users other than root.

Summing up the discussion above the following gives a brief idea about the challenges that need to be overcome:

- A useful secure operating system should make it easier to write secure applications.
- There is a need for more flexible permission model. The models present today are either too simple or too restrictive.
- The issue here is that, no commercial operating system is secure enough.
- There will always be buggy code, but the trick is to build an application and an operating system that will mostly restrict attacks and will protect the important assets of the system.

Security is not only an issue with the operating systems in desktops and laptops; the operating systems of tablets and cell-phones also have the same security issues but these issues in phones are the most critical because if an attacker gets into the operating system of a phone, the attacker may get access to the personal data (viz. contacts, messages, etc) of the victim. And moreover the user base of these smaller devices like smart-phones and tablets in increasing at an alarming rate and the amount of data sharing between these devices is far more than that between computers.

## 3. MEMORY MANAGEMENT

Managing the system memory is a very important function of an operating system. Hence the success of any operating system also depends to some extent on how well the operating system manages the system memory. There have been numerous mechanisms that have been researched upon and implemented in this area of operating system development. Today, an operating system has to execute tasks on a huge amount of data but in the early days the catch was that to operate on data, it had to be present in the primary memory and primary memory cannot be as much as the secondary memory. So the researchers and developers started finding alternate ways of storage and execution of data. During this time came a concept called paging.

In operating systems, paging is one of the memory management schemes by which the system can store and retrieve data from the secondary storage for use in the main memory. In this scheme, the operating system retrieves data from secondary storage in same size blocks called pages. The main function of paging is performed when a program tries to access pages that are not currently mapped to the RAM. This situation is known as a page fault. When page fault occurs, an operating system has to perform the following tasks:

- Determine the location of data in auxiliary storage.
- Obtain an empty page frame in RAM to use as a container for data.
- Load the requested data into the available page frame.
- Update the page table to show the new data.
- Return control to the program, transparently retrying the instruction that caused the page fault.

Until there is not enough RAM to store all the data needed, the process of obtaining an empty page frame does not involve removing another page from RAM. If all page frames are non-empty, obtaining an empty page frame requires choosing a page frame containing data to empty. If the data in that page frame has been modified since it was read into RAM, it must be written back to its location in secondary storage before being freed; otherwise, the contents of the page's page frame in RAM are the same as the contents of the page in secondary storage, so it does not need to be written back to secondary storage. If a reference is then made to that page, a page fault will occur, and an empty page frame must be obtained and the contents of the page in secondary storage again read into that page frame. Efficient paging systems must determine the page frame to empty by choosing one that is least likely to be needed within a short time. There are various page replacement algorithms that try to do this. Most operating systems use some approximation of the least recently used (LRU) page replacement algorithm (the LRU itself cannot be implemented on the current hardware) or a working set-based algorithm.

Paging is a very important feature for memory management and is made use of by most of the commercially available operating systems. For example, consider paging in Windows. Almost all memories in windows can be paged out to disks. This is where page file comes into play; its where most pages are placed when they are not resident in the physical memory. However, not everything gets written into page files, they get written to specific mapped files. Better than that, the pages only get written if they have been modified. If they have not been altered since they were read from the file, windows doesn't have to write the pages back out; it can just discard them. If it ever needs the pages again, they can be safely re-read from the files. Although paging is a very efficient mechanism yet challenges still exist in this area, that need to be overcome if the performance of the system has to be increased.

Operating systems today have taken paging to the next level, by allowing sharing of pages between different processes. This technique has an important advantage, that is, it avoids duplication of pages for multiple processes. Or in other words, if pages were not shared between processes, then each process would have had to acquire its own copy of a page that is being used by another process. Hence by allowing sharing of pages, the execution time of instructions goes down, in turn making the operating system run faster. This memory sharing is useful, especially in low-memory systems, but the current technique present for sharing of pages, has its limitations; major one being that the operating system only shares memory that corresponds to memory mapped files. That is because this is the only time that the operating system knows that pages are identical. For regular data there is no page sharing.

A new scheme for page sharing is going to be implemented by vendors. Here, the system will periodically scan memory, and when it finds two pages that are identical, it will share them, reducing the memory usage. If a process then tries to modify the shared page, it will be given its own private copy, ending the sharing. This mechanism will have a huge effect on virtualization. When virtualizing, the same operating system may be running multiple times, meaning that the same executable files are loaded several times over. So the traditional memory-mapped file approach to memory sharing cannot kick in here. Each virtual operating system is loading its own files from its own disk image. This is where memory de duplication is useful; it can see that the pages are all identical, and hence it can allow sharing even between virtual machines.

This is another technique that is used by some operating systems (Mac OSX) for memory management. As per this method, when the operating system needs memory it will push something that isn't currently being used into a swap file for temporary storage. When it needs access to that data again, it will read the data from the swap file and back into memory. In a sense this can create unlimited memory, but it is significantly slower since it is limited by the speed of the hard disk, versus the near immediacy of reading data from RAM. Even this mechanism has a flaw. For example, consider that processes A, B, C are to be executed one after the other wherein A and C need same resources but B needs totally different resources. Another assumption here is that there is no memory left in the RAM. So here once process A is finished, process B will have to run, but since B needs different resources and resources of A are not required anymore for now, they are shifted into swap file and resources for B are loaded in place of that. Now when C is to be executed, again the resources that had been shifted to swap file has to be shifted back to the RAM. So here we see how redundant swapping of data takes place and this results in slow processing speed.

The following points sum up the areas of concern for an operating system to obtain more efficient memory management:

- The operating systems today use some approximation of the LRU (least recently used) algorithm as the LRU itself has not been completely implemented on any present machine.
- To increase responsiveness, paging systems must employ better strategies to predict which page will be needed soon. Such systems will attempt to load pages into main memory preemptively, before a program references them.
- Operating systems will need better methods of page sharing, such that page sharing for regular data and not only for memory-mapped data can be achieved.
- If swapping mechanism is to be used for memory management, then proper measures need to be taken to avoid redundant sharing of data as much as possible.

## 4. MULTIPROCESSOR PROGRAMMING

Now a days usage of more than one processors in a computing system has become a common occurrence. Operating systems should have efficient mechanism to support more than one processors and the ability to schedule tasks between them. There are many variants of this basic theme and the definition of multiprocessing may vary with context.

In a multiprocessing system, all CPUs may be equal, or some may be reserved for special purposes. A combination of hardware and OS software design considerations determine the symmetry or lack of it in a given system. For example, hardware or software considerations may require that only one CPU respond to all hardware interrupts, whereas all other work in the system may be distributed equally among CPUs; or execution of kernel-mode code may be restricted to only one processor at a time whereas user-mode code may be executed in any combination of processors. Multiprocessing systems are often easier to design if such restrictions are imposed, but they tend to be less efficient than systems in which all CPUs are utilized. Systems that treat all CPUs equally are called Symmetric Multiprocessing Systems (SMP). In systems where CPUs are not equal, system resources may be divided in a number of ways including Asymmetric Multiprocessing Systems (ASMP), Non-Uniform Memory Access (NUMA) multiprocessing systems and Clustered Multiprocessing Systems.

In computing, SMP involves a multiprocessor computer architecture where two or more identical processors can connect to a single shared main memory. Most common multiprocessor systems today use SMP architecture. In case of multi-core processors, the SMP architecture applies to the cores, treating them as separate processors. SMP systems allow any processor to work on any task no matter where the data for that task is located in the memory. With proper OS support SMP systems can easily move tasks between processes to balance the workload efficiently.

Asymmetric multiprocessing varies greatly from the standard processing model that we see in the personal computers today. Due to the complexity and unique nature of this architecture it was not adopted by many vendors during a brief stint. While SMP treats all of the processing elements in the system identically, an ASMP system assigns certain tasks only to certain processors. Although hardware level ASMP may not be in use, the idea and logical process is still commonly used in applications that are multiprocessor intensive. Unlike SMP applications which run there threads on multiple processors, ASMP application will run on one processor but outsource smaller tasks to other processors. The operating systems may also make use of ASMP architecture for critical tasks like the tasks that may make use of system files. Operating systems can dedicate one processor called the Master processor to implementation of tasks required on the system files while smaller related tasks may be delegated to other processors called the Slave Processors. Although the basic architecture will still be SMP yet for critical tasks the ASMP architecture may be used.

Modern CPUs operate considerably faster than the main memory they use. In the early days of computing and data processing the CPU generally ran slower than its memory. The performance lines crossed in the 1960s with the advent of high speed computing. Since then, CPUs increasingly "starved for data", have had to stall while they wait for memory accesses to complete. Limiting the amount of memory access provides the key to extracting high performance from a modern day computer. For commodity processors this means installing an ever increasing amount of high speed cache memory and very sophisticated algorithm to avoid cache misses. But dramatic increases in size of the operating systems make the problem considerably worse. Now a system can starve several processors at the same time, notably because only one processor can access memory at a time. NUMA attempts to address this problem by providing separate memory for each processor, avoiding performance hit when several processors attempt to address the same memory. Of course not all data ends up confined to a single task, which means that more than one processor may require the same data. To handle these cases, NUMA systems include additional hardware or software to move data between banks. This operation slows the processors attached to those banks, so the overall speed increase due to NUMA depends heavily on the exact nature of tasks that are running. This architecture can substantially increase the performance but for that there has to be proper hardware and the operating system must provide some mechanism to efficiently schedule the access to multiple processor memory. If NUMA architecture is implemented successfully both in the hardware and in the OS level then it could go a long way in speeding up processing with multiple processors.

The following points highlight the areas of research and development for efficient multiprocessor programming by modern day operating systems:

- Although most of the operating systems today use SMP architecture yet with proper operating system support SMP systems can move tasks between processors more freely and thus balance the workload effectively.
- Operating Systems can implement a hybrid of SMP and ASMP architectures wherein, while all the tasks can be delegated using SMP architecture, the tasks that make use of system files can make use of ASMP architecture to implement that part.
- NUMA architecture can be seriously looked upon during future operating system design such that a way to integrate this architecture into the system is reached. If this happens, it could go a long way in speeding up the processing with multiple processors.

**5. User Interface Design**

Let's start this segment by saying that the future is mobile, and there is a little dispute about this. Desktop machines will only be used for very heavy specialized purposes, the same way trucks are used today. Most of the people will just own fast-enough mobile, portable devices rather than desktops. Basically, anything bigger than a 5" screen will be too much to carry. In time carrying a 10" tablet will seem no different than the way we today feel about 1981's businessman, carrying around the Osborn 1.

In human-computer interaction WIMP stands for "Windows, Icons, Menus and Pointers"; denoting a style of interaction using these elements. WIMPs are systems where a window will run a self-contained program isolated within that window from other programs running at the same time. Icons act as shortcuts to the actions to be performed by the system, Menus are text or icon based selection systems to select and execute programs or sub-programs and finally, Pointer is an on-screen symbol that represents the movement of a physical device to allow the user to select elements on an output device such as a monitor. User interfaces base on WIMP are very good at abstracting workspaces, documents and their actions. Their basic representations as rectangular regions on 2D flat screens make them a good fit for system programmers. Generality makes them very suitable for multi-tasking work environments.

However researchers consider this to be a sign of stagnation in user interface design as the path of least resistance forces developers to follow a particular way of interaction. There are applications for which WIMP is not well suited, they argue, and the lack of technical support increases difficulty for development of interfaces not based on WIMP style. This includes any application requiring devices that provide continuous input signals, showing 3D models or simply portraying an interaction for which there are no defined standard widgets. WIMPs are usually pixel-hungry. So given limited screen real-estate, they can distract attention from the task at hand. Thus custom interfaces can better encapsulate workspaces, action and other objects from specific complex tasks. Interface based on these considerations now called post-WIMP are making their way to the general public.

The following points highlight the issues of WIMP from a touch-GUI perspective:

- *Pointers*- We cannot have any sort of pointer indicators when touching the screen.
- *Windows*- From a touch perspective, Windows are almost completely useless. Moving, resizing, minimizing, maximizing, closing are all things that are just plain too hard to do and only create extra overhead on the small display screen.
- *Menus*- Traditional window menus are super useful things to have in computers. But that said, they are tiny and hard to manage with fingers and if one bumps up the size of the fonts more, he might as well throw away a third of the screen real-estate.

In short, there are just too many fundamental issues with the WIMP to just tweak. It's not a matter of size, weight, power or probability of the devices that matter- the core under-printing of WIMP based interface is just incompatible with touch usability and everything is going to need to be re-written from ground-up. Moreover, since the researchers and developers now are talking about one operating system for all the devices, hence this transition from the traditional WIMP will soon be needed for all major operating systems.

Putting together the entire discussion the following points highlight the challenges in development of the user interface for operating systems:

- Since the devices are getting smaller and smaller, a way has to be found to port the traditional WIMP applications for these smaller devices.
- Major changes will have to be made to the operating systems and the applications such that they can run in both our traditional desktops and the smaller touch devices (smart-phones and tablets) such that we may achieve a level where one operating system could be used in all the devices. To fully achieve this we will also have to find a way as to how the WIMP applications could be used with modern touch devices.

## 6. CONCLUSION

As the user awareness of technology is increasing so is there expectations. Hence although operating systems have progressed a lot, yet still there is a lot of ground to cover in this field. Operating systems research is a very vast field and the reason for this is mostly because the hardware is becoming stronger and faster by the day and hence there is a race for the operating systems to keep up. The key issues pointed out in this paper if addressed, will make our computation even more wonderful than the present.

## 7. REFERENCES

[1] Galen C. Hunt, James R. Larus, David Tarditi and Ted Wobber. Brand New OS Research: Challenges and Opportunities, UNISEX.

[2] Schneider, F.B. Enforceable Security Policies. ACM Transactions on Information and System Security(TISSEC).

[3] Abraham Silberschatz, Peter Bear Galvin and Gary Gagne. Operating System Concepts.

[4] C. Kaner and D.L. Pels. Bad Software: What To Do When Software Fails.

[5] http://www.wikipedia.org/

[6] Buck, T. Foley, D. Horn, J. Sugerman, K. Fatahalian, L. M. Houston and P. Hanrahan. Brooks for GPUs: stream computing on graphics hardware. Proc. Of the 2004 SIGGRAPH Conference, pp. 777-786, 2004.

[7] S. Chaki, S. K. Rajamani and J. Rehof. Types as Models: Model Checking Message Passing Programs. Proc of the 29$^{th}$ ACM Symposium on Principles of Programming Languages, pp. 45-57, 2002.

[8] J. DeTreville, Making System Configuration more Declarative. Proc. Of Hot OSX: The 10$^{th}$ Workshop on Hot Topics in Operating Systems, June 2005.

**Mr. Plawan Kumar Rath** is currently perusing his Bachelor's degree in BMS Institute of Technology, Department of CSE, Bangalore. He is in the final year of his Bachelor's degree and is scheduled to be associated with IBM India Software Labs after completion of his bachelor's degree in summer of 2012. He has a keen interest in research and development mostly in the field of operating systems and system softwares.

**Mr. Anil G. N.** is currently working as Associate Professor in the Department of CSE in BMS Institute of Technology. He has a vast teaching experience in the field of Computer Sciences. He has published a number of papers in National and International Conferences and Journals and won numerous accolades.